\documentclass[11pt, a4paper]{article}
\pdfoutput = 1
\usepackage[height=22cm,width=16cm, centering]{geometry}

\usepackage{setspace}
\usepackage[sort&compress, numbers, merge]{natbib}

\usepackage{amsmath, amssymb, mathrsfs,  comment}
\usepackage{ifpdf, xcolor}

\ifpdf        
  \usepackage{graphicx}     
  \usepackage[bookmarksopen,colorlinks=true, linkcolor=link_col,
  citecolor=cite_col, urlcolor=url_col,linktocpage=false]{hyperref}
\else     
  \usepackage[dvipdfmx]{graphicx}     
  \usepackage[dvipdfmx,bookmarksopen,colorlinks=true,linkcolor=link_col,
  citecolor=cite_col ,urlcolor=url_col,linktocpage=false]{hyperref}
\fi

\usepackage{multicol}
\definecolor{link_col}{rgb}{0.0, 0, 0.8}
\definecolor{cite_col}{rgb}{0.6, 0, 0.3}
\definecolor{url_col}{rgb}{0.6, 0, 0.3}

\def\Mpl{M_{\rm P}}
\def \psiL{\psi^{\ell}}

\def \psiT{\psi^t}
\def \vecpsiT{\vec{\psi}^{t}}

\begin{document}

\begin{titlepage}
\begin{center}
\leavevmode \\

{\small 
\hfill IPMU17-0008\\
\hfill KIAS-P17003\\
\hfill  UT-17-01}

\noindent
\vskip 1.5 cm
{\huge Gravitino Problem in Minimal Supergravity Inflation}

\vskip 0.8 cm

{\Large Fuminori Hasegawa$^1$, Kyohei Mukaida$^2$, Kazunori Nakayama$^3$,\\ Takahiro Terada$^4$, and Yusuke Yamada$^5$}

\vskip 1. cm

{\textit {\small
$^1$Institute for Cosmic Ray Research, The University of Tokyo, Kashiwa, Chiba 277-8582, Japan\\
$^2$Kavli IPMU (WPI), UTIAS, The University of Tokyo, Kashiwa, Chiba 277-8583, Japan\\
$^3$Department of Physics, Faculty of Science, The University of Tokyo,\\  Bunkyo-ku, Tokyo 133-0033, Japan\\
$^4$School of Physics, Korea Institute for Advanced Study (KIAS), 
Seoul 02455, Republic of Korea\\
$^5$Stanford Institute for Theoretical Physics and Department of Physics,\\
Stanford University, Stanford, CA 94305, U.S.A.}}

\vskip 1.6 cm

\begin{abstract}
{\normalsize 
We study non-thermal gravitino production in the minimal supergravity inflation.
In this minimal model utilizing orthogonal nilpotent superfields, the particle spectrum includes only graviton, gravitino, inflaton, and goldstino.
We find that a substantial fraction of the cosmic energy density can be transferred to the longitudinal gravitino due to non-trivial change of its sound speed.  This implies either a  breakdown of the effective theory after inflation or a serious gravitino problem.
}
\end{abstract}

\end{center}

\end{titlepage}

\section{Introduction}

Supersymmetry (SUSY) and supergravity are well-motivated candidates for particle physics beyond the Standard Model. 
Although SUSY has been escaping from collider searches as well as (in)direct detection experiments so far, it may well be relevant in the early universe.
Hence, it is useful to constrain models from cosmological considerations.  A classical example is the gravitino problem~\cite{Pagels:1981ke, Weinberg:1982zq, Khlopov:1984pf,Ellis:1984eq}.
Gravitino, being the superpartner of graviton, has extremely weak interactions and is long-lived, which affects late-time cosmology, and hence its abundance is tightly constrained~\cite{Ellis:1984er, Moroi:1993mb, Moroi:1995fs, Jedamzik:2004er, Kawasaki:2004yh, Kawasaki:2004qu, Jedamzik:2006xz, Kawasaki:2008qe}.
Gravitino is produced thermally~\cite{Bolz:2000fu,Pradler:2006qh,Rychkov:2007uq, Ellis:2015jpg} or non-thermally~\cite{Kallosh:1999jj, Giudice:1999yt, Giudice:1999am, Kallosh:2000ve, Nilles:2001ry, Nilles:2001fg, Ema:2016oxl, Endo:2006zj,Nakamura:2006uc,Kawasaki:2006gs, Asaka:2006bv,Dine:2006ii,Endo:2006tf,Kawasaki:2006hm,Endo:2007ih,Endo:2007sz} in the cosmological history.

Another central issue in SUSY cosmology is realization of inflation in supergravity.
While SUSY is useful to stabilize the flatness of the inflaton potential against radiative corrections,
there are many scalar fields in supergravity in general, and the whole scalar dynamics tends to be complicated.
For example, one has to ensure that the scalar potential is stable in all the directions in the scalar manifold, 
and that the reheating is successful without producing dangerous relics too much.

Recently, there has been a much progress in this respect, utilizing constrained superfields~\cite{Rocek:1978nb, Komargodski:2009rz, Antoniadis:2014oya, Ferrara:2014kva, Kallosh:2014via, Aoki:2014pna, Dall'Agata:2014oka, Kallosh:2014hxa, Scalisi:2015qga, Ferrara:2015tyn, Dall'Agata:2015lek, Kallosh:2016hcm, Dall'Agata:2016yof, McDonough:2016der}, which signal the non-linear realization of SUSY~\cite{Volkov:1973ix,Ivanov:1978mx}.
One can remove components of a supermultiplet from the particle spectrum, which can be interpreted as decoupling of heavy particles.
In particular, the effective field theory (EFT) for SUSY inflation with the minimal degrees of freedom has been constructed, in which there are only graviton, gravitino, (real) inflaton and goldstino~\cite{Kahn:2015mla, Ferrara:2015tyn, Carrasco:2015iij, Delacretaz:2016nhw} explaining also the low-energy SUSY breaking and dark energy~\cite{Ferrara:2015tyn, Carrasco:2015iij}.
This is achieved using orthogonal nilpotent superfields~\cite{Komargodski:2009rz, Kahn:2015mla, Ferrara:2015tyn, Carrasco:2015iij}. It is worth mentioning that the model with constrained superfields would be realized as a low-energy EFT of superstring~\cite{Ferrara:2014kva, Vercnocke:2016fbt, Kallosh:2016aep}.

In this Letter, we study non-thermal gravitino production in the (p)reheating stage after inflation in the model with orthogonal nilpontent superfields.
It was envisaged~\cite{Ferrara:2015tyn, Carrasco:2015iij} that the analysis of gravitino production is significantly simplified due to the fact that the only fermion in the spectrum is gravitino after the super-Higgs mechanism.
In fact, the analysis of gravitino production with multiple superfields is notoriously complicated since gravitino mixes with a fermion in a time-dependent fashion~\cite{Kallosh:2000ve,Nilles:2001ry,Nilles:2001fg}, and analytic understanding for realistic cases has been achieved only recently~\cite{Ema:2016oxl}.
We find, however, that the longitudinal gravitino generically behaves quite differently from the standard (unconstrained) cases, and it can be produced extravagantly depending on the parameters and cutoff of the model.
This may lead to a cosmological catastrophe, and the validity of the inflation in supergravity with the minimal degrees of freedom should be carefully reconsidered.
We also discuss possibilities to circumvent the problem at the end of the Letter.

\section{Model}
We consider the minimal supergravity inflation model whose general K\"{a}hler and superpotentials can be written as follows~\cite{Ferrara:2015tyn, Carrasco:2015iij}:
\begin{align}
K= & \bar{S}S + k(\Phi, \bar{\Phi}), \\
W= & S f(\Phi) + g(\Phi),
\end{align}
with orthogonal nilpotent superfields $S$ and $\Phi$ satisfying the constraints
\begin{align}
S^2 = & 0 & \text{and}&  &  
S (\Phi - \bar{\Phi}) = & 0. \label{constraints}
\end{align}
That is, $S$ is nilpotent and $\Phi - \bar{\Phi}$ is orthogonal to $S$. 
This implies the nilpotency condition with degree three, $(\Phi-\bar{\Phi})^3 = 0$~\cite{Komargodski:2009rz}. 
Under these constraints, we can express the ``sgoldstino'' $S$, inflatino $\widetilde{\Phi}$, sinflaton ${\rm Im}\,\Phi$, and $F$-term of $\Phi$ in terms of the ``goldstino'' $\widetilde{S}$. The true goldstino in the cosmological background is a combination of $\widetilde{S}$ and $\widetilde{\Phi}$~\cite{Kallosh:2000ve}. 
 The goldstino and all these dependent fields vanish in the unitary gauge $\widetilde{S}=0$~\cite{Ferrara:2015tyn}, which we take in the following analysis.

The scalar potential is different~\cite{Ferrara:2015tyn, Carrasco:2015iij} from that of the textbook-level supergravity, 
$
V = e^{k / \Mpl^2} ( |f|^2 - 3 |g|^2 \Mpl^{-2}),
$
 where $\Mpl$ is the reduced Planck mass. 
The canonically normalized inflaton is denoted by $\phi$, and its mass by $m_{\phi}$.

The Lagrangians of the canonically normalized transverse and longitudinal gravitinos, $\vecpsiT$ and $\psiL$, in the conformal metric $ds^2 = a(\eta)^2 (-d\eta^2+d\vec{x} ^2)$ and in the unitary gauge $\widetilde{S}=0$ are~\cite{Ferrara:2015tyn, Ema:2016oxl}
\begin{align}
	\mathcal L_t = & -\frac{1}{2}\overline{\vecpsiT} \cdot \left[ \gamma^0 \partial_0 +  \left( \vec\gamma\cdot\vec \nabla \right) + a m_{3/2} \right] \vecpsiT,
	\label{Ltrans} \\
	\mathcal L_\ell = & - \frac{1}{2}\overline{\psiL} \left[ 
		\gamma^0\partial_0 - \widehat {c}_{3/2} \left( \vec\gamma\cdot\vec \nabla \right) 
		+ a\widehat{m}_{3/2}
	\right]\psiL,  
	\label{grav_kin}	
\end{align}
where the transverse gravitino mass is $m_{3/2} = g(\Phi) \Mpl^{-2}$, 
the speed-of-sound and mass parameters for the longitudinal mode are given by
\begin{align}
	\widehat {c}_{3/2} \equiv & \frac{p_{\rm SB} - \gamma^0 p_{W}}{\rho_{\rm SB}}, \\
	\widehat{m}_{3/2} \equiv &
	\frac{3Hp_{W} + m_{3/2}(\rho_{\rm SB}+3p_{\rm SB})}{2\rho_{\rm SB}},\label{hatm3/2}
\end{align}
and the SUSY breaking contribution to the energy density $\rho$ and the pressure $p$ are defined as
\begin{align}
	\rho_{\rm SB} \equiv \rho + 3 m_{3/2}^2 \Mpl^2,~~~~~p_{\rm SB} \equiv p - 3 m_{3/2}^2 \Mpl^2, 
\end{align}
with $p_{W} \equiv 2\dot m_{3/2} \Mpl^2$ measuring the time-derivative of the gravitino mass. 
A dot denotes a derivative with respect to the standard time $t$ ($\partial_0 = a \partial_t$).

\section{Gravitino production}

In our setup, the sound speed of the longitudinal gravitino $|\widehat{c}_{3/2}|$ is generically not equal to one, $|\widehat{c}_{3/2}|\neq 1$. 
This change of sound speed is in contrast to the standard (unconstrained) cases~\cite{Nilles:2001fg, Ema:2016oxl}.
This non-relativistic feature was reported in Refs.~\cite{Kahn:2015mla, Ferrara:2015tyn} (see also Refs.~\cite{Lebedev:1989rz, Kratzert:2003cr, Hoyos:2012dh, Benakli:2014bpa, Benakli:2013ava}). 
Below, we show that the time variation of this quantity produces many longitudinal gravitinos.

First, note that $p_W$ is negligible during inflation, during the radiation dominant epoch, and in the present universe.  Then, the sound speed parameter $\widehat{c}_{3/2}$ is nothing but the equation-of-state parameter of the cosmic component that breaks SUSY, $\widehat{c}_{3/2} \simeq p_{\text{SB}}/\rho_{\text{SB}} \equiv w_{\text{SB}}$.
As we will see shortly, the change of the sign of this quantity produces/annihilates the longitudinal gravitino.
This implies the mechanism is insensitive to the details of the thermal or expansion history of the universe after inflation.
We do not have to even assume the existence of the period of the inflaton coherent oscillation.

To see the production of the longitudinal gravitino, we expand the field in the Dirac representation as
\begin{align}
\psiL =\sum_h \int \frac{d^3 k}{(2\pi)^{3/2}} e^{i \vec k \cdot \vec x} \begin{pmatrix} u^+_{\vec k, h}(t) \\ u^-_{\vec k, h}(t) \end{pmatrix} \otimes \xi_{\vec k,h} \, \hat{b}_{\vec k, h}+ \text{H.c.},
\end{align}
where $\xi_{\vec k, h}$ is the normalized eigenvector of helicity $h$,  $(\vec \sigma \cdot \vec{k})\xi_{\vec k,h}=h|\vec k| \xi_{\vec k,h}$, and $\hat{b}_{\vec k, h}$ is an annihilation operator.
(An analogous expression holds for the transverse mode.)
For our purpose, we only have to consider $u^+_{\vec k, h}$ with $h$ either $1$ or $-1$, so we omit the superscript and the subscripts.
The relevant mode equation for the longitudinal gravitino is
\begin{align}
&c_{3/2} u'' - c'_{3/2} u'+ c_{3/2}\widetilde{\omega}_{k}^2 u = 0, \label{modeEq+}  \\
&\widetilde{\omega}_{k}^2  \equiv a^2 \omega_k^2 - i c_{3/2}(a\widehat{m}_{3/2}/c_{3/2})', \label{omegaTilde}
\end{align}
where a prime denotes the conformal time derivative, $c_{3/2} \equiv (p_{\text{SB}}+i p_W)/\rho_{\text{SB}}$, and $\omega_k\equiv \sqrt{\widehat{m}_{3/2}^2 + (c_{3/2} k /a)^2}$ can be interpreted as the energy of one gravitino particle.
The effects of Hubble expansion have been taken into account although they are implicit in the conformal coordinates.

We first consider the simplest possibility of constant $g(\Phi)=m_{3/2} \Mpl^{2}$, and discuss more general cases later.
In this case, $p_W$ vanishes, and $c_{3/2}=c_{3/2}^*=w_{\text{SB}}$.
Let us first consider the $m_{3/2} \rightarrow 0$ limit.
The solution of Eq.~\eqref{modeEq+} is
\begin{align}
u(\eta ) = \frac{1}{\sqrt{2}} \exp \left [ ik\int_0^\eta \text{d}\eta' c_{3/2}(\eta') \right ].  \label{mode_solution}
\end{align}
This satisfies the vacuum initial condition, $u(0) = 1/\sqrt{2}$ and $u'(0)= - i \omega_k (0) u(0)$.
In deriving Eq.~\eqref{mode_solution}, we have not assumed any specific form of $c_{3/2}(t)$,  and this implies robustness of the mechanism against back-reaction effects. 
The Bogoliubov coefficients (the coefficients of the positive and negative frequency modes) switch when the speed-of-sound parameter $c_{3/2}$ crosses zero, signaling particle production.
In fact, the phase space distribution function is obtained as follows,
\begin{align}
f_{3/2}(\vec k ; t) \equiv & \frac{1}{2 \omega_k (t)} \left(  2 \text{Im}\, \left( u^{+ *}_{\vec k} (t) \dot{u}^{+}_{\vec k} (t) \right) - \widehat{m}_{3/2}(t) \right) + \frac{1}{2} \nonumber \\
=& \frac{1}{2} \left( 1 +  \text{sgn} \left( c_{3/2} (t) \right) \right). \label{PhaseSpaceDist}
\end{align}
Thus, particle production or annihilation occurs when $c_{3/2}$ changes its sign. 
Note that this is independent of the momentum $k$, and high-momentum modes with $k \gg m_{\phi}$ can also be produced.
Restoring the (small) gravitino mass $m_{3/2}$ smears the sharp edge of $f_{3/2}(t)$ and modulates the oscillation in the longer time scale $m_{3/2}^{-1}$, both of which would not affect our order-of-magnitude estimations and conclusions significantly.

We need to discuss the momentum cutoff, otherwise the solution Eq.~\eqref{PhaseSpaceDist} itself predicts particle production with infinitely high momenta.
There are two possibilities:  (A) This EFT is invalid for the discussion on the above mechanism of gravitino production since it involves too high momentum and energy.  One has to use a UV completed theory to discuss the (p)reheating process. We do not discuss this option further. 
(B) The EFT is valid, and the momentum cutoff is given by the UV cutoff scale $\Lambda$ of the EFT up to some coefficients which we assume to be order one.
\footnote{
On the dimensional ground, the momentum cutoff would be $\Lambda/c_{3/2}$ with $\Lambda$ defined as the energy cutoff.  Since the particle production occurs when $c_{3/2}$ changes its sign, the momentum cutoff may be enhanced.  Since $c_{3/2}$ is a time-dependent quantity, it is not clear whether we can substitute $c_{3/2}\rightarrow 0$ into the formula, which induces divergence.  The $\Lambda$ (as well as $\Lambda_{\text{UB}}$) itself may be proportional to $c_{3/2}$, but this makes the possibility (A) more likely.  Thus, the following estimate is conservative (a lower bound) regarding the abundance of gravitinos.
} 
  The cutoff may be in general time- (or field-)dependent and should be higher than the Hubble scale $H$ and lower than the expected unitarity bound $\Lambda_{\text{UB}}\equiv ( \Mpl^2 (H^2 + m_{3/2}^2) )^{1/4}$~\cite{Casalbuoni:1988sx, Kallosh:2000ve, Dall'Agata:2014oka, Kahn:2015mla, Ferrara:2015tyn, Carrasco:2015iij, Delacretaz:2016nhw}.
The longitudinal gravitino number density becomes
\begin{align}
a^3 n_{3/2}  \sim 2 \int _0 ^\Lambda dk \frac{4\pi k^2}{(2\pi )^3} f_{3/2}(\vec k ; t) \sim \Lambda^3. 
\end{align}
Similarly, the energy density of the created particles is $a^4 \rho_{3/2} \sim \Lambda^4$.
We assume this is sufficiently smaller than the total energy density $\rho=3H^2\Mpl^2 \sim \Lambda_{\text{UB}}^4$ so that we can safely neglect back-reaction effects.  For later use, we define the ratio $r \equiv (\Lambda^4 / \rho)^{3/4}$ which measures how much fraction of (the 3/4 power of) the total energy density is converted to that of gravitinos.

Since $\widehat{c}_{3/2}\simeq -1$ during inflation and at the late time ($H\ll m_{3/2}$), the zero-crossing of the sound speed occurs even times.  One may wonder that all the created gravitinos are annihilated, but generically this complete cancellation does not occur because of cosmic expansion and redshift of modes.  For instance, if the equation-of-state parameter satisfies $0<w_{\text{SB}}<1/3$ before the final transition at $H \sim m_{3/2}$, high momentum modes near the cutoff scale are created.  Also, if there is the period of inflaton coherent oscillation, $f_{3/2}(k)$ has an oscillating feature averaging to $1/2$.  These situations also generate $\rho_{3/2} \sim \Lambda^4$ at the final zero-crossing at $H \sim m_{3/2}$.

If the reheating is completed at $H \gtrsim m_{3/2}$, the gravitino yield, $Y_{3/2} \equiv n_{3/2}/s$ where $s$ is the entropy density, becomes
\begin{align}
Y_{3/2} \sim r,  \label{Y32EarlyDecay}
\end{align}
which is fixed at $H\sim m_{3/2}$ due to the above mechanism.  
For a wide range of $r$, this is clearly a cosmological disaster because it exceeds the constraint on the dark matter abundance~\cite{Ade:2015xua}
\begin{align}
Y_{3/2} \lesssim 5 \times 10^{-15} \times \left( \frac{10^5 \, \text{GeV}}{m_{\text{LSP}}} \right), \label{DMconstraint}
\end{align}
if the $R$-parity is conserved assuming that one gravitino produces one lightest supersymmetric particle (LSP) by its decay and LSPs do not efficiently annihilate thereafter.
There is also the constraint $Y_{3/2} \lesssim 10^{-13} \sim10^{-16}$ for the GeV--TeV range gravitino from big-bang nucleosynthesis (BBN)~\cite{Moroi:1995fs, Jedamzik:2004er, Kawasaki:2004yh, Kawasaki:2004qu, Jedamzik:2006xz, Kawasaki:2008qe}.
It is possible that the gravitino decays above the LSP freeze-out temperature for
\begin{align}
m_{3/2} \gtrsim 2 \times 10^7 \,\text{GeV} \left( \frac{m_{\text{LSP}}}{10^3\, \text{GeV}} \right)^{2/3}.
\end{align}
In this case the LSP abundance crucially depends on the properties of the LSP, although typically it tends to be too abundant for such a heavy gravitino scenario.

On the other hand, if the reheating is completed at $H\lesssim m_{3/2}$, gravitino is diluted after $H \sim m_{3/2}$ until the reheating.
For simplicity, let us consider the inflaton coherent oscillation dominance which behaves as non-relativistic matter.
The yield is estimated as
\begin{align}
Y_{3/2}  \sim 10^{-3} \, r \left( \frac{T_{\text{R}}}{10^9 \, \text{GeV} } \right) \left( \frac{10^5 \,\text{GeV}}{m_{3/2}} \right)^{1/2}, \label{Y32LateDecay}
\end{align}
where $T_{\text{R}}$ is the reheating temperature.
This is still a quite huge abundance compared to the above constraints.  
In addition, there is a contribution of the gravitino produced by the perturbative inflaton decay, which we will mention soon.

As we have seen, the case of constant $g(\Phi)$ produces an enormous amount of gravitinos, so let us turn to more general cases. When we turn on small $g_\Phi$, this explosive particle production is still effective, but there is a cutoff on the possible highest momentum.  This can be obtained from the condition that the second term in the equation of motion~\eqref{modeEq+} is always subdominant: 
$k_{\text{max}} \sim (\Mpl^2 H^2 m_{\phi})/|g_\Phi|^2$. 
 We expect that the actual cutoff is the lowest of this and $\Lambda$.

This mechanism of particle creation by sound speed change can be understood also by the standard technique of preheating analyses.
After a $\gamma^0$-dependent phase rotation of the gravitino, $\psiL = e^{\theta \gamma^0} \tilde{\psi}^{\ell}$, the sound-speed parameter becomes a positive semidefinite number and does not change its sign.
Instead, the time derivative of the phase rotation parameter appears as an effective mass term contribution,
\begin{align}
\mathcal L_\ell = & - \frac{1}{2}\overline{\tilde{\psi}^{\ell}} \left[ 
		\gamma^0\partial_0 -  |c_{3/2}| \vec\gamma\cdot\vec \nabla 
		+ a(\widehat{m}_{3/2} - \theta')
	\right]\tilde{\psi}^{\ell}, \label{LlongTransformed}
\end{align}
where $\tan 2\theta = - p_W / p_{\text{SB}}$.
The effective mass oscillates like spikes, and gravitinos with a broad (almost entire) band of wave numbers are excited. 
For example, in the constant $g(\Phi)$ case, the effective mass is a delta function, $\theta' =  \pi \delta(p_{\text{SB}}) p'_{\text{SB}} / 2$ leading to an extremely non-adiabatic oscillation. 
In more general cases with nonzero $|g_{\Phi}|$, we reproduce the same cutoff $k_{\text{max}} \sim (\Mpl^2 H^2 m_{\phi})/|g_\Phi|^2$  from the condition of non-adiabaticity, $\dot{\omega}_{k_{\text{max}}} \sim \omega_{k_{\text{max}}}^2$.

A large $g_\Phi$, however, does not effectively relax our situation of the violent gravitino production.
If it is large when the inflaton decays perturbatively to reheat the universe, 
the longitudinal gravitino is again copiously produced by the inflaton decay provided that the decay occurs when $H \lesssim m_{3/2}$. 
The decay rate is calculated by the technique used in Ref.~\cite{Ema:2016oxl} from the Lagrangian~\eqref{LlongTransformed}, 
\begin{align}
\Gamma (\phi \rightarrow \psiL \psiL)  = \frac{|g_{\Phi}|^2 m_{\phi}^5}{288\pi m_{3/2}^4 \Mpl^4}, \label{DecayRate}
\end{align}
which reproduces the perturbative decay rate in Refs.~\cite{Endo:2006zj, Nakamura:2006uc}. 
We end up with
\begin{align}
Y_{3/2} \simeq & 2 \times 10^{-2} \left( \frac{m_{\phi}}{10^{13}\, \text{GeV}} \right)^4 \left( \frac{10^{9}\, \text{GeV}}{T_{\text{R}}} \right) \nonumber \\
& \qquad \times \left( \frac{10^{5}\, \text{GeV}}{m_{3/2}} \right)^{4}  \left( \frac{|g_\Phi|}{(10^9 \, \text{GeV})^2} \right)^2. \label{Y32reheating}
\end{align}
Naively, we need $|g_\Phi| \sim |f| \sim H \Mpl$ at the oscillation period to avoid the excessive gravitino production.  Extrapolating this relation to the reheating epoch, we obtain an upper bound on the reheating temperature through $T_{\text{R}} \sim \sqrt{|g_\Phi|}$.

The above results imply that $g(\Phi)$ cannot be a constant, but its first derivative must be suppressed.
Therefore, we consider the case with a sufficiently large second derivative $g_{\Phi\Phi}|_{\phi=0}$.
In this case, both the transverse and longitudinal gravitinos are produced by their oscillating $\phi^2$-dependent mass.  
We estimate the production rate as~\cite{Ema:2016oxl}
\begin{align}
\Gamma (\phi \phi \rightarrow \psiT \psiT \text{ or } \psiL \psiL) \simeq \frac{3 H^2 |g_{\Phi\Phi}|^2}{2 \pi m_{\phi} \Mpl^2},
\end{align}
up to order one uncertainty, leading to 
\begin{align}
Y_{3/2} \simeq  & 6 \times 10^{-16}  \left( \frac{T_{\text{R}}}{10^{9}\, \text{GeV}} \right) \left(  \frac{H_{\text{inf}}}{10^{13}\, \text{GeV}} \right)  \nonumber \\
 & \qquad \times \left( \frac{10^{13} \, \text{GeV}}{m_{\phi}} \right)^2 \left( \frac{|g_{\Phi\Phi}|}{10^{13}\, \text{GeV}} \right)^2, \label{Y32Annihilation}
\end{align}
where $H_{\text{inf}}$ is the Hubble parameter just after inflation, again up to order one uncertainty. 
Thus, we have obtained a constraint also on the magnitude of $g_{\Phi\Phi}$. 
This constraint itself is relatively not so severe, but one should keep in mind that just increasing $|g_{\Phi}|$ or $|g_{\Phi\Phi}|$ does not necessarily ensure that we can neglect the contribution of Eq.~\eqref{Y32EarlyDecay} or \eqref{Y32LateDecay}.
We also need to care about the effect on the inflationary trajectory since $g(\Phi)$ gives a negative contribution to the scalar potential.

\section{Discussion}
In this Letter, we studied non-thermal gravitino production at the (p)reheating stage of the minimal supergravity inflation model, namely with the orthogonal nilpotent superfields.
This model was well-motivated both from theoretical/phenomenological viewpoints involving application of non-linear realization of SUSY to cosmology and from the perspective of EFT of SUSY inflation.
It can describe not only inflation but also the low-energy SUSY breaking and the cosmological constant with the minimal particle contents.
One special feature of this model is that the goldstino is described by the same system throughout the whole history of the universe from the inflationary era to the present. 
Gravitino production depends on the functional form of $m_{3/2}(\Phi)=g(\Phi) \Mpl^{-2}$, and a non-negligible amount of gravitinos is generically produced [Eqs.~\eqref{Y32EarlyDecay} or \eqref{Y32LateDecay}, \eqref{Y32reheating}, and \eqref{Y32Annihilation}] provided that the EFT is applicable. 
For the simplest case with constant $g(\Phi)$, the longitudinal gravitino is explosively produced in the wide momentum range up to the UV cutoff scale $\Lambda$ of the model [Eq.~\eqref{Y32EarlyDecay} or \eqref{Y32LateDecay}].  
In conclusion, we end up with either the breakdown of the EFT after inflation or the serious gravitino problem for simple or generic choices of the gravitino mass function $g(\Phi)$.

The cause of such an explosive production is the change of the sound speed of the longitudinal gravitino.
As shown in Refs.~\cite{Nilles:2001fg, Ema:2016oxl}, this does not occur in the standard cases without constrained superfields or higher derivatives taking into account all the mixings of gravitino with other fermions. 
The analysis in each UV completed model of the minimal supergravity inflation is beyond the scope of this Letter, and should be studied elsewhere.

There is an exceptional case: $g_\Phi=f$, which leads to $|c_{3/2}|=1$.
The reason why this case does not lead to sound speed change can be understood from the fact that the low-energy limit of unconstrained single-superfield models of inflation and SUSY breaking~\cite{Izawa:2007qa, AlvarezGaume:2010rt, Achucarro:2012hg, Ketov:2014qha, Schmitz:2016kyr, Ferrara:2016vzg} with non-minimal K\"{a}hler potential can be described by orthogonal nilpotent superfields with $g_\Phi=f$. See Appendix E of Ref.~\cite{Komargodski:2009rz}.
Although the production through sound speed change does not happen in this case, a huge amount of gravitinos is produced by the inflaton decay: $g_{\Phi}= \sqrt{3} m_{3/2}\Mpl$ in Eqs.~\eqref{DecayRate} and \eqref{Y32reheating}.

We have skipped the discussion on the dynamics of the coupled system of inflaton and the longitudinal gravitino.  The details of back-reactions depend on precisely how much fraction of the energy density is transferred to the gravitino, which we just parametrize as $r=\Lambda^3/\rho^{3/4}$.  However, it is impossible to derive it within the EFT framework because it involves the cutoff $\Lambda$ of the EFT. 
Nevertheless, we believe our qualitative argument is sufficient for order-of-magnitude estimations because the production mechanism depends only on the sign change of the speed-of-sound parameter $\widehat{c}_{3/2}$ or the equation-of-state parameter $w_{\text{SB}}$.
At least, our results serve as an important caution against the minimal supergravity inflation.

Once accepting the EFT below the UV cutoff, it is difficult to avoid the explosive gravitino production since non-minimal forms of $g(\Phi)$ may also lead to a significant abundance of gravitino [Eqs.~\eqref{Y32reheating} and \eqref{Y32Annihilation}].
However, it does not necessarily mean a cosmological disaster.
One can assume a sufficiently heavy gravitino not to affect the BBN or a large entropy release/$R$-parity violation to dilute the produced gravitinos/LSPs, although one has to give up a simple/minimal setup from the model building viewpoint. 
Hence, our result calls for a careful study whether a given UV theory is reduced to the effective theory with the orthogonal nilpotent superfields.

Taking our results seriously, there may be several implications for the modification of the minimal supergravity inflation.
For example, introduction of non-minimal ingredients such as inflatino can evade the change of the sound speed.
Alternatively, we can add additional low-energy SUSY breaking fields.  This allows the produced gravitinos morph into $\widetilde{S}$ if the additional fields dominantly break SUSY at late time.  It will be less harmful than gravitino, but detailed cosmological consequences depend on its coupling to other fields. 
Also, it will be interesting to study how gravitino production is modified when we remove the orthogonality constraint in Eq.~\eqref{constraints}.

Finally, we note that the same mechanism of particle production (for either bosons or fermions) via sound speed change is possible in a more general non-SUSY context. 
For the scalar case, this is known as the gradient instability, see \text{e.g.} Refs.~\cite{Ohashi:2012wf,Ema:2015oaa} and also Ref.~\cite{Giblin:2017qjp}.
 Sound speed oscillation can be induced by derivative couplings with inflaton, which are totally allowed from the EFT point of view. 
This effect is, however, expected to end quickly with moderate choices of parameters when the leading term dominates in the derivative expansion.
The longitudinal gravitino is special in this point because of its characteristic speed of sound.
We will study these points and gravitino production in more details in future~\cite{future}.

\section*{Acknowledgments}
The authors are grateful to Renata Kallosh and Andrei Linde for useful comments on their manuscript. 
FH thank Yutaro Shoji for useful discussion. KM, KN, and TT thank Yohei Ema for discussion.
TT is grateful to Kenji Kadota, Tomoki Nosaka, and Yasuhiro Yamamoto for discussions. YY would like to thank Renata Kallosh for useful discussion about orthogonal superfields and Lorentz violation.
This work is supported in part by National Research Foundation
of Korea (NRF) Research Grant NRF-2015R1A2A1A05001869
[TT], JSPS Grant-in-Aid for Scientific Research on Scientific Research
A (JP26247042), Young Scientists B (JP26800121) and Innovative
Areas (JP26104009 and JP15H05888) [KN], Grant-in-Aid
for JSPS Research Fellow JP15J07034, JSPS Research Fellowships for
Young Scientists, World Premier International Research Center Initiative
(WPI Initiative), MEXT, Japan [KM], SITP and the NSF grant
PHY-1316699 [YY].

\small

\bibliographystyle{utphys}
\bibliography{orthogonal_nilpotent}

\end{document}